\documentclass[prd,showpacs]{revtex4}
\usepackage{amssymb}
\usepackage{amsmath}
\usepackage{graphicx}
\usepackage{epstopdf}

\allowdisplaybreaks

\begin{document}

\title{New general DBI action, its solution to the paradox of the conversion
of kinetic and potential energy in
equal rights and their true applications to inflationary
cosmology}
\author{Xiaokun Yang$^{1}$}
\email{xkyang@emails.bjut.edu.cn}
\author{Wu-Long Xu $^{1}$}
\email{wlxu@emails.bjut.edu.cn}
\author{Yong-Chang Huang $^{1,2}$}
\email{ychuang@bjut.edu.cn}
\affiliation{$^{1}$ Institute of Theoretical Physics, Beijing University of Technology,
Beijing, 100022, China\\
$^{2}$Institute of Theoretical Physics, Jiangxi Normal University, Nanchang 330022, China}
\date{\today }

\begin{abstract}
The Dirac-Born-Infeld (DBI) field theory in string theory is important and can provide the
field of the universe's inflation. At the same time, it provides a causal
mechanism for generating the original density perturbation, thereby providing
the necessary density perturbation for existing the dense and sparse matter
distributions of the universe. However, there is the paradox of
the conversion of potential energy and kinetic energy in equal rights in string theory. Therefore,
we give a new general DBI action, which enables the kinetic energy and potential
energy in the action to be converted each other in equal rights, i.e.,
solving the paradox. Therefore, we deduce a new general DBI action, introduce it
into inflationary cosmology to calculate various inflation parameters,
further calculate the scalar perturbation spectrum and the tensor-scalar
ratio, which are compared with Planck + WMAP9 + BAO data, the power spectrum predicted by the new general DBI
inflation theory satisfies the CMB Experiment constraints, i.e., is
consistent with the current theories and experimental observations. Consequently, the theory of this paper conforms to current experiments and is supplying the current
theories, and also a new way of explaining the inflation of the universe.

\end{abstract}

\maketitle

\section{Introduction}

Current cosmology has two important discoveries, one is Hubble expansion,
and the other is cosmic microwave background radiation. These two
discoveries tell us that our universe is expanding, and it is uniform and
flat~\cite%
{Spergel:2006hy,Alabidi:2006qa,Seljak:2006bg,Kinney:2006qm,Cole:2005sx,Barger:2003ym}
. Therefore the production mechanism of such a flat universe was proposed.
For explaining the flatness of the universe, the horizon of the universe and
the problem of magnetic monopoles, the most mainstream theory now is the
theory of cosmic inflation. It means that the universe experienced a rapid
acceleration expansion in the early stage of the creation, and the causal
connection area rapidly expanded into a large area for forming an observable
universe, and the non-flatness was smoothed. Moreover, the theory of cosmic
inflation can provide the original perturbation, thus giving birth seeds to
the galaxy structure. The cosmic inflation is usually realized by scalar
fields~\cite%
{Guth:1980zm,Linde:1981mu,Fayyazuddin:1993eb,Debono:2021vzc,Lidsey:1995np,Linde:2005ht,Riotto:2010jd}
. We call this scalar field as inflation field.

In the very early stage of the universe, the initial value of the potential
energy $V\left( \varphi \right) $ of the inflation field $\varphi $ was very
large, and the potential energy curve was very flat. This potential energy
provides the energy of the universe's inflation. With the evolution of time
and the change of the inflation field, the potential energy slowly
decreases, and the scaling factor of the universe increases almost
exponentially in this process, resulting in the universe inflation that we
predict.

The general theory of cosmic inflation does not give the source of the
scalar field. A major contribution of string theory is that the DBI field
theory produced by string theory can provide the field of the universe's
inflation~\cite{Linde:1981mu,Albrecht:1982wi,Dvali:1998pa, Kachru:2003sx}.
This theory shows that the inflation of the universe is related to the
motion of D-brane in the 6-dimensional compact submanifold of spacetime. It
is the motion of a certain string that causes the scalar field to be
generated, and it restricts the motion of the scalar field. This is an
effective field theory explanation.

This explanation is based on string theory and is quite consistent in terms
of the construction mechanism. When we integrate other dimensions to 4
dimensions, the DBI action will naturally be deduced. String theory also
gives the behavior of cosmic inflation, so people can make general
cosmological predictions based on a set of given background parameters~\cite%
{Chen:2006nt}. This makes the brane universe inflation a good theory for
exploring the universe's inflation. Therefore, it is possible to establish
the universe evolution theory that satisfies various parameters and is
consistent with the basic theory. There has been a lot of work in studying
the inflation of the universe and understanding the inflation field~\cite%
{Liu:2009vx,Kinney:2007ag,Tolley:2008na,Peiris:2007gz,ArmendarizPicon:1999rj, Garriga:1999vw,Shandera:2006ax}
.

The DBI action can constructs the inflationary cosmology ~\cite%
{Alishahiha:2004eh,Silverstein:2003hf}, which considers that ~$Ads_{5}$~
space is related to a "throat"-shaped area. Then in this region, the radial
coordinates of the moving $D_{3}-$ brane are taken as a scalar field in our
universe. With the deformation of ~$Ads_{5}$~ space, this throat-like area
will also be deformed, then the moving $D_{3}-$ brane will be compressed,
causing it to change accordingly~\cite{Silverstein:2003hf}. The change of
the radial coordinate of the $D_{3}-$ brane, that is, the movement of this
scalar field, is affected by the distortion coefficient of the throat~\cite%
{Huang:2006eha,Chen:2006nt}.

When inflationary cosmology solves the problem of flatness, monopoles and
the difficulty of horizon, it is natural that due to the quantum perturbation
in the inflation process, it will produce seeds of large-scale structure for
the universe. There are two sources of perturbation. One is the perturbation
caused by the stretching of spacetime itself in inflation. This perturbation
is embodied as a gravitational wave. We call it adiabatic perturbation or
tensor perturbation~\cite{Maldacena:2002vr}, the other comes from the quantum
perturbation caused by the scalar field that produces inflation, we call it
scalar perturbation. The source of tensor perturbation is essentially
different from scalar perturbation.

However, none of the current DBI theories considers the paradox, for example
the literature ~\cite{ArmendarizPicon:1999rj,Bean:2007hc,Lidsey:2007gq}, of
the conversion of potential energy and kinetic energy in equal rights. The
problem, in other words, the conversion of potential energy and kinetic
energy with each other in these actions can only be carried out under the
first-order approximation, and the potential energy can not be completely
converted to kinetic energy under high-order. Therefore, we want to give a
new general DBI action, which enables the kinetic energy and potential
energy in the action to be converted each other in equal rights. Since we
used the conformal transformation when obtaining the new general DBI action,
the potential energy $V$ will be naturally generated from this
transformation, instead of manually adding the potential energy term like
the current general DBI action. This makes our theory be consistent.

The new general DBI action has the advantage of equal rights of kinetic
energy and potential energy, so whether it can meet with the existing
observational data is the main research point. After we deduce the new
general DBI action by string theory, we introduce it into inflationary
cosmology to calculate various inflation parameters, and calculate the
scalar perturbation spectrum and tensor-scalar ratio, further compare them
with Planck + WMAP9 + BAO data to verify the correctness of the new general
DBI action. Whether it conforms to the observational data and whether it
will violate the prediction of the standard DBI action, then we need to
verify whether the new general DBI action can be driven by inflation and
produce the vast universe we see now. And lots of DBI theories and their applications to different physics systems have been very well given out, e.g., see refs.~\cite{Garrett:2021,Pran:2019,Zhewei:2019,Tanguy:2020,Ignatios:2020,Shuntaro:2017,Sjoerd:2016}.

In the second section, we derive a new general DBI Lagrangian for inflation.
This Lagrangian contains various possible interaction potentials.
Furthermore, we naturally derive the determinant of the induced metric, and
then compare the new general DBI Lagrangian with the current general DBI
Lagrangian. In the third section, we obtained the cosmological equations and
cosmological parameters deduced from the new general Lagrangian, and
calculate the slow-rolling parameters and e-ford number. In the fourth
section, we calculate the power spectrum of the original perturbation under
the linear perturbation, and calculate the form of the tensor-scalar ratio.
In order to reflect the relationship between the original scalar perturbation
and the inflation field, we use perturbation to directly calculate the
magnitude of the perturbation amplitude to verify the true role of the
inflation field in it. In the fifth section, we have selected different
background fields and potential functions, and obtained the functional
relationship diagrams between the tensor-scalar ratio index of the original
scalar perturbation, and compared it with the background of the Planck + WMAP9 + BAO joint data. In the two cases of $N=50$ and $N=60$
, the tensor-scalar ratio is reflected by different parameter groups, thus
their correct properties are proved. The final section is conclusion.

\section{ new general DBI action}

In this section, we will start with string theory, consider the conformal
transformation to derive a new general DBI action, and compare it with the
current general DBI action.

Here we deduce a new general Lagrangian, and find that the determinant of
the metric can naturally include momentum energy and potential energy in
equal rights.

The current general DBI action is in the form of ~\cite{Kinney:2006,Kahya:2008,String Theory:2006}:

\begin{equation}
\mathcal{L\text{=}}-f^{-1}\left(\varphi\right)\sqrt{1+f\left(\varphi
\right)g^{\mu\nu}\partial_{\mu}\varphi\partial_{\nu}\varphi}
-V\left(\varphi\right)+f^{-1}\left(\varphi\right)  \label{L1}
\end{equation}

Obviously, in such a Lagrangian, the kinetic energy $\partial _{\mu }\varphi
\partial _{\nu }\varphi $ and the potential energy $V\left( \varphi \right) $
have the paradox of unequal rights. Therefore, we need to consider a new
general DBI action and solve the paradox for the current general DBI action.
When establishing this new general DBI action, we consider that the field
that promotes the inflation of the universe exists in a $\left( 1+3\right) $
dimensional spacetime and a 6-dimensional curly spacetime. And regarding the
$\left( 1+3\right) $ dimensional world body as a particle migrating along
the radial $r$ and passing through the 6-dimensional curly spacetime~\cite%
{Albrecht:1982wi,Firouzjahi:2005dh, Spalinski:2007kt,Bardeen:1980kt}. Then
the corresponding line element can be expressed as ~\cite{Liu:2009vx}:

\begin{equation}
ds_{10}^{2}=h^{2}\left(r\right)ds_{4}^{2}+h^{-2}\left(r\right)
\left(dr^{2}+r^{2}ds_{y_{5}}^{2}\right)
\end{equation}

Take the metrics of $ds_{4}^{2}$ and $ds_{y_{5}}^{2}$ as $a_{\mu \nu }\left(
x\right) $ and $a_{ab}\left( y\right) $ respectively and do conformal
transformation on them:

\begin{equation}
a^{\prime
}_{\mu\nu}\left(x\right)=W_{1}\left(r\right)a_{\mu\nu}\left(x\right)
\label{a2-3}
\end{equation}
\begin{equation}
a^{\prime }_{ab}\left(y\right)=W_{1}\left(r\right)a_{ab}\left(y\right)
\label{a2-4}
\end{equation}

Then, the metric matrixes can be written as:

\begin{equation}
G=\left[
\begin{array}{ccc}
h^{2}\left(r\right)a_{\mu\nu}\left(x\right) &  &  \\
& h^{-2}\left(r\right) &  \\
&  & h^{-2}\left(r\right)r^{2}a_{ab}\left(y\right)%
\end{array}
\right]
\end{equation}

\begin{equation}
G^{\prime }=\left[
\begin{array}{ccc}
h^{2}\left(r\right)W_{1}\left(r\right)a_{\mu\nu}\left(x\right) &  &  \\
& h^{-2}\left(r\right) &  \\
&  & h^{-2}\left(r\right)W_{1}\left(r\right)r^{2}a_{ab}\left(y\right)%
\end{array}
\right]
\end{equation}

Under the conformal transformations~\eqref{a2-3} and~\eqref{a2-4}, the
induced world volumes elements are invariant, namely:

\begin{equation}
\sqrt{-detG}d^{4}x=\sqrt{-detG^{\prime }}d^{4}x^{\prime }  \label{G}
\end{equation}

Then, integrating eq.~\eqref{G}, we deduce a general action:

\begin{equation}
S=\chi\int d^{4}x\sqrt{-detG}=\chi\int d^{4}x^{\prime }\sqrt{-detG^{\prime }}
\label{intG}
\end{equation}

\noindent where $\chi$ is an invariant parameter with  dimension so that the product of $\chi$ and the world body integral is the general action, and the general action~\eqref{intG} is invariant under the conformal
transformation.

Therefore, in string theory and M theory, the induced metric in our
four-dimensional spacetime is:

\begin{equation}
G_{\alpha\beta}=h^{2}\left(r\right)W_{1}\left(r\right)a_{\mu\nu}\left(x
\right)\frac{\partial x^{\mu}}{\partial x^{\alpha}}\frac{\partial x^{\nu}}{
\partial x^{\beta}}+h^{-2}\left(r\right)\frac{\partial r}{\partial
x^{\alpha} }\frac{\partial r}{\partial x^{\beta}}+h^{-2}\left(r\right)W_{1}%
\left(r \right)r^{2}a_{ab}\left(y\right)\frac{\partial x^{a}}{\partial
x^{\alpha}} \frac{\partial x^{b}}{\partial x^{\beta}}
\end{equation}

Since $y$ has nothing to do with $x$ in $\left( 1+3\right) $ dimensional
spacetime, where~$\mu ,\nu ,\alpha ,\beta \left( =0,1,2,3\right) $, $%
a,b\left( =5,6,7,8,9\right) $, $x^{4}=r$~. Thus, the third term in~$%
G_{\alpha \beta }$~satisfies

\begin{equation}
h^{-2}\left(r\right)W_{1}\left(r\right)r^{2}a_{ab}\left(y\right)\frac{
\partial x^{a}}{\partial x^{\alpha}}\frac{\partial x^{b}}{\partial x^{\beta}}
=0
\end{equation}

Define $W_{1}\left(r\right)=1+W_{11}\left(r\right)$, where $%
W_{11}\left(r\right)$ is also a general function of $r$ , then the inducted
metric can be written as:

\begin{equation}
G_{\alpha\beta}=h^{2}\left(r\right)g_{\alpha\beta}+h^{2}\left(r\right)W_{11}
\left(r\right)g_{\alpha\beta}+h^{-2}\left(r\right)\frac{\partial r}{\partial
x^{\alpha}}\frac{\partial r}{\partial x^{\beta}}
\end{equation}

\noindent $W_{11}\left(r\right)$ is a general function.

Therefore, the new general DBI action on the $D_{3}-$ brane can be written
as:

\begin{equation}
\begin{aligned} S_{DBIL}= & -T\intop
d^{4}\sigma\sqrt{-det\left[h^{2}\left(r\right)g_{\alpha\beta}+h^{2}\left(r
\right)W_{11}\left(r\right)g_{\alpha\beta}+h^{-2}\left(r\right)\frac{
\partial r}{\partial x^{\alpha}}\frac{\partial r}{\partial
x^{\beta}}\right]}\\ = & -T\intop d^{4}\sigma
h^{4}\left(r\right)\sqrt{-detg_{\alpha\lambda}}\sqrt{det\left[\delta_{
\beta}^{\lambda}\left(1+W_{11}\left(r\right)\right)+h^{-4}\left(r\right)g^{
\lambda\tau}\frac{\partial r}{\partial x^{\tau}}\frac{\partial r}{\partial
x^{\beta}}\right]} \end{aligned}
\end{equation}

We can get the string tension as:

\begin{equation}
T_{p_{d}}=\frac{1}{g_{s}\left(2\pi\right)^{d}\left(\alpha^{\prime
}\right)^{2}}
\end{equation}

So, for the $D_{3}-$ brane,

\begin{equation}
T_{p_{3}}=\frac{1}{g_{s}\left(2\pi\right)^{3}\left(\alpha^{\prime
}\right)^{2}}
\end{equation}

\noindent It is a function of string coupling strength $g_{s}$, so we can
define inflationary scalar field $\varphi =r\sqrt{T_{p_{3}}}$.

Then, our DBI action can be written as:

\begin{equation}
S_{DBIL}=-T_{p_{3}}\intop d^{4}\sigma h^{4}\left(\varphi\right)\sqrt{
-detg_{\alpha\lambda}}\sqrt{det\left[\delta_{\beta}^{\lambda}\left(1+W_{11}
\left(\varphi\right)\right)+h^{-4}\left(\varphi\right)g^{\lambda\tau}
\partial_{\tau}\varphi\partial_{\beta}\varphi\right]}
\end{equation}

Further, the action of DBI is written as:

\begin{equation}
\begin{aligned} S_{DBIL} & =-T_{p_{3}}\intop d^{4}\sigma
h^{4}\left(\varphi\right)\sqrt{-detg_{\alpha\lambda}}\sqrt{det\left[\delta_{
\beta}^{\lambda}\left(1+W_{11}\left(\varphi\right)\right)+h^{-4}\left(
\varphi\right)g^{\lambda\tau}\partial_{\tau}\varphi\partial_{\beta}\varphi
\right]}\\ & =-\intop d^{4}\sigma
f^{-1}\left(\varphi\right)\sqrt{-detg_{\alpha\lambda}}\sqrt{det\left[
\delta_{\beta}^{\lambda}\left(1+W_{11}\left(\varphi\right)\right)+f\left(
\varphi\right)g^{\lambda\tau}\partial_{\tau}\varphi\partial_{\beta}\varphi
\right]} \end{aligned}  \label{lashi}
\end{equation}

\noindent where

\begin{equation}
f\left(\varphi\right)=\frac{1}{T_{p_{3}}h^{4}\left(\varphi\right)}
\end{equation}

When taking $\sqrt{-detg_{\alpha \lambda }}$ to the invariant volume
element of the integral ~\eqref{lashi} and adding an integral scalar
term $T_{p_{3}}\intop d^{4}\sigma h^{4}\left( \varphi \right) \sqrt{
-detg_{\alpha \lambda }}=\intop d^{4}\sigma \sqrt{-detg_{\alpha \lambda }}
f^{-1}\left( \varphi \right) $ to ~\eqref{lashi}, we finally obtain a new
general DBI Lagrangian.

\begin{equation}
\mathcal{L}{}_{DBIL}=-f^{-1}\left(\varphi\right)\sqrt{det\left[
\delta_{\beta}^{\lambda}\left(1+W_{11}\left(\varphi\right)\right)+f\left(
\varphi\right)g^{\lambda\tau}\partial_{\tau}\varphi\partial_{\beta}\varphi %
\right]}+f^{-1}\left(\varphi\right)  \label{xin-}
\end{equation}

The $f\left(\varphi\right)$ can be regarded as the background scalar field
of the universe.

Because in the earliest period of the universe, all kinds of matter fields
did not appear in the universe, but at this time the gravitational field and
scalar field had already begun to act, so other types of type IIB fields did
not appear in the action. Since $W_{11}\left(\varphi\right)$ is an arbitrary
function, we define $W_{11}\left(\varphi\right)=f\left(\varphi\right)KV
\left(\varphi\right)$ ($K$ is any parameter). Under linear approximation,
the eq.~\eqref{xin-} is rewritten as:

\begin{equation}
\mathcal{L}{}_{DBIL}=-f^{-1}\left(\varphi\right)\sqrt{1+f\left(\varphi%
\right) \left[g^{\alpha\beta}\partial_{\alpha}\varphi\partial_{\beta}%
\varphi+K V\left(\varphi\right)\right]}+f^{-1}\left(\varphi\right)
\label{xin}
\end{equation}

The eq.~\eqref{xin} uses a parameter to ensure that the kinetic energy item $%
g^{\alpha \beta }\partial _{\alpha }\varphi \partial _{\beta }\varphi $ is
in the same equal position as the potential energy. This is the basic
Physical requirements and general actual physical systems. Because if the
system cannot naturally contain kinetic energy and potential energy, the
system is not a general actual physical system. Therefore, it can be found
that the conformal transformation is a very useful transformation, which has
an important physical meaning relative to the curl factor $h$, which just
keeps the kinetic energy and potential energy always in equal rights each
other.

The eq.~\eqref{xin} can be linearly approximated to the current general DBI
Lagrangian. Taking the linear approximation of Lagrangian, we have:

\begin{equation}
\mathcal{L}{}_{DBIL}=-\frac{1}{2}g^{\alpha\beta}\partial_{\alpha}\varphi
\partial_{\beta}\varphi-\frac{1}{2}K V\left(\varphi\right)  \label{xinxian}
\end{equation}

Refer to the current general DBI Lagrangian~\eqref{L1} given by ~\cite%
{Kinney:2006,Kahya:2008,String Theory:2006}.The potential energy $V$ in this
Lagrangian~\eqref{L1} looks as if it was directly added manually, and the
kinetic energy and the potential energy are not in equal rights, because

\begin{equation}
\mathcal{L}{}_{DBI}=\left(-\frac{1}{2}g^{\alpha\beta}\partial_{\alpha}
\varphi\partial_{\beta}\varphi-V\left(\varphi\right)\right)+\frac{1}{8}
f\left(\varphi\right)\left(g^{\alpha\beta}\partial_{\alpha}\varphi\partial_{\beta}\varphi
\right)^{2}-\frac{1}{16}f^{2}\left(\varphi\right)\left(g^{\alpha\beta}\partial_{\alpha}\varphi
\partial_{\beta}\varphi\right)^{3}\cdots  \label{yuan}
\end{equation}

Do a linear approximation to this Lagrangian, one can get:

\begin{equation}
\mathcal{L}{}_{DBI}=-\frac{1}{2}g^{\alpha\beta}\partial_{\alpha}\varphi
\partial_{\beta}\varphi-V\left(\varphi\right)  \label{xian}
\end{equation}

If in a physical system, the kinetic energy and the potential energy cannot
be transformed each other in equal rights, anyone can believe that this
is not a real physical situation, and this new general DBI action can give
the transformation of kinetic energy and potential energy in equal rights.

The comparing formula ~\eqref{xinxian} with eq.~\eqref{xian}, if we take $%
K=2 $, the eq.~\eqref{xinxian} will be simplified to ~\eqref{xian}. This
shows that Lagrangian ~\eqref{L1} is only a special case of our new general
DBI Lagrangian.

Note that eq.~\eqref{xin} is more general than eq.~\eqref{L1}, because eq.~ %
\eqref{xin} shows that potential energy can be converted to kinetic energy
in equal rights, and the potential energy naturally emerges from the
Lagrangian, without manual adding to. Eq.~\eqref{xian} is only a linear
approximation of the special DBI action, and the eq.~\eqref{xin} is the new
general Lagrangian of the DBI action. The potential energy in the formula of
~\eqref{L1} cannot be directly converted into kinetic energy in equal
rights, which destroys the basic physical principles. The new general DBI
Lagrangian~\eqref{xin} is different from previous work. Eq.~\eqref{xin}
reveals the basic physical meaning of the potential energy generated by $%
D_{3}-$brane. This section solves the paradox of conversion of potential
energy and kinetic energy in equal rights.

\section{Inflationary Cosmology and Inflationary Parameters}

In this section, we will use the new general DBI action to generate the
inflationary cosmological action, get the density and momentum driven by the
inflation field in the early universe, and calculate the inflation parameter
and e-ford number.

We apply this action to the dynamics of the inflationary universe~\cite%
{Nozari:2013wua}:

\begin{equation}
S=\intop dx^{4}\sqrt{-g}\left[\frac{1}{\kappa^{2}}R-f^{-1}\left(\varphi
\right)\sqrt{1+f\left(\varphi\right)\left[g^{\alpha\beta}\partial_{\alpha}
\varphi\partial_{\beta}\varphi+K V\left(\varphi\right)\right]}
+f^{-1}\left(\varphi\right)\right]  \label{eq:zuoyongliang}
\end{equation}

\noindent where $R$ is a Ricci scalar, then the Einstein equation obtained
from this action is:

\begin{equation}
G_{\mu\nu}=\kappa^{2}T_{\mu\nu}
\end{equation}

$T_{\mu\nu}$ comes from the DBI action:

\begin{equation}
\begin{aligned} T_{\mu\nu}= &
\frac{2\partial\left[-f^{-1}\left(\varphi\right)\sqrt{1+f\left(\varphi
\right)\left[g^{\alpha\beta}\partial_{\alpha}\varphi\partial_{\beta}
\varphi+K V\left(\varphi\right)\right]}\right]}{\partial g^{\mu\nu}}\\ &
-g_{\mu\nu}\left[-f^{-1}\left(\varphi\right)\sqrt{1+f\left(\varphi\right)
\left[g^{\alpha\beta}\partial_{\alpha}\varphi\partial_{\beta}\varphi+K
V\left(\varphi\right)\right]}+f^{-1}\left(\varphi\right)\right] \end{aligned}
\label{T}
\end{equation}

Therefore, different metric needs to be selected to find the corresponding
Einstein field equation.

Select FRW metric:

\begin{equation}
ds^{2}=dt^{2}-a^{2}\left(t\right)\left(\delta_{ij}+k\frac{x_{i}x_{j}}{
1-kr^{2}}\right)dx^{i}dx^{j}
\end{equation}

\noindent where $k=-1,0,1$. Using eq.~\eqref{T}, the density and momentum of
matter can be obtained as:

\begin{equation}
\rho=-\frac{\dot{\varphi}^{2}}{\sqrt{1+f\dot{\varphi}^{2}+K fV}}
-f^{-1}+f^{-1}\sqrt{1+f\dot{\varphi}^{2}+K fV}
\end{equation}

\begin{equation}
p=\frac{\partial_{j}\varphi\partial^{j}\varphi}{3\sqrt{1+f\dot{\varphi}
^{2}+K fV}}+f^{-1}-f^{-1}\sqrt{1+f\dot{\varphi}^{2}+K fV}
\end{equation}

Solving the Einstein equation, we get the Friedmann equation:

\begin{equation}
H^{2}=\frac{\kappa^{2}}{3}\left(-\frac{\dot{\varphi}^{2}}{\sqrt{1+f\dot{
\varphi}^{2}+K fV}}-f^{-1}+f^{-1}\sqrt{1+f\dot{\varphi}^{2}+K fV}\right)
\end{equation}

Doing variation on the action ~\eqref{eq:zuoyongliang}, we get the equation
of motion:

\begin{equation}
\begin{aligned} & \frac{\ddot{\varphi}}{\left(1+f\dot{\varphi}^{2}+K
fV\right)^{\frac{3}{2}}}+\frac{3H\dot{\varphi}}{\left(1+f\dot{\varphi}^{2}+K
fV\right)^{\frac{1}{2}}}\\ = &
-f'f^{-2}+f'f^{-2}\left(1+f\dot{\varphi}^{2}+K
fV\right)^{\frac{1}{2}}-\frac{1}{2}\frac{f'f^{-1}\dot{\varphi}^{2}+K
f'f^{-1}V+K V'}{\left(1-f\dot{\varphi}^{2}+K fV\right)^{\frac{1}{2}}}
\end{aligned}  \label{eq:yundongfangcheng}
\end{equation}

If we consider the slow roll approximation $\dot{\varphi}^{2}\ll1$ and $%
\ddot{\varphi}\ll\left|3H\dot{\varphi}\right|$, the energy density of the
DBI field and the equation of motion will become:

\begin{equation}
\rho=-f^{-1}+f^{-1}\sqrt{1+K fV}
\end{equation}

\begin{equation}
H^{2}=\frac{\kappa^{2}}{3}\frac{1}{f}\left(\sqrt{1+K fV}-1\right)
\end{equation}

\begin{equation}
3H\dot{\varphi}=-f^{\prime -2}\sqrt{1+K fV}+f^{\prime -2}+\frac{1}{2}K
f^{\prime -1}V-\frac{1}{2}K V^{\prime }
\end{equation}

Then we can get the slow roll parameter $\epsilon\equiv-\frac{\dot{H}}{H^{2}}
$, $\eta\equiv-\frac{\ddot{H}}{H\dot {H}}$, namely:

\begin{equation}
\begin{aligned}\epsilon\equiv-\frac{\dot{H}}{H^{2}}= &
\frac{\kappa^{2}}{18H^{4}}(1+K
fV)^{-\frac{1}{2}}\left[f^{-2}f'-f^{-2}f'\sqrt{1+K fV}+\frac{1}{2}K
f^{-1}f'V-\frac{1}{2}K V'\right]^{2}\\ = & \frac{\kappa^{2}}{18H^{4}}(1+K
fV)^{-\frac{1}{2}}Y^{2} \end{aligned}
\end{equation}

\begin{equation}
\eta=-\frac{Y}{3H^{2}}\left(\frac{2Y^{\prime }}{Y}-\frac{Y}{H^{2}}+\frac{1}{%
2 }\frac{K f^{\prime }V+K fV^{\prime }}{1+K fV}\right)
\end{equation}

\noindent where

\begin{equation}
Y=-f^{\prime -2}\sqrt{1+K fV}+f^{\prime -2}+\frac{1}{2}K f^{\prime -1}V-
\frac{1}{2}K V^{\prime }
\end{equation}

The inflation occurs under the conditions of $\left\{ \epsilon,\eta\right\}
<1$; once these slow-rolling parameters reach $1$, the inflation phase ends.
The e-ford number during the inflation period is defined as:

\begin{equation}
N=\int_{t_{hc}}^{t_{f}}Hdt
\end{equation}

For our model, the e-ford number during slow scrolling can be expressed as:

\begin{equation}
N=\int_{\varphi_{hc}}^{\varphi_{f}}\frac{3H^{2}}{3H\dot{\varphi}}
d\varphi=\kappa^{2}\int_{\varphi_{hc}}^{\varphi_{f}}\frac{1-\sqrt{1+K fV}}{
f^{\prime -1}\sqrt{1+K fV}-f^{\prime -1}+\frac{1}{2}K f^{\prime }V-\frac{1}{%
2 }K fV^{\prime }}d\varphi
\end{equation}

\noindent where $\varphi_{hc}$ represents the field value when the observed
universe scale crosses the Hubble horizon during inflation, and $\varphi_{f}$
is the field value when the universe exits the inflation phase.

\section{Inflation Perturbation}

In this section, we will study the theory of linear perturbations by the new
general DBI action.

In many different methods, according to the choice of the metric that
characterizes the cosmic perturbation, we choose the longitudinal metric. The
scalar metric perturbation of the FRW background is given by the literature ~%
\cite{Bardeen:1980kt,Mukhanov:1990me,Bertschinger:1993xt}:

\begin{equation}
ds^{2}=\left(1+2\Phi\right)dt^{2}-a^{2}\left(t\right)\left(1-2\Psi\right)
\delta_{ij}dx^{i}dx^{j}  \label{rao}
\end{equation}

\noindent where $a\left( t\right) $ is the scale factor of the universe,
metric perturbations $\Phi =\Phi \left( x,t\right) $ and $\Psi =\Psi \left(
x,t\right) $ are gauge invariant variables. All perturbations behave like
plane waves $e^{ikx}$ in space, where $k$ is the wave number. Through
Einstein's field equations, any perturbation of the metric will cause
perturbations of the energy-momentum tensor. For the perturbed metric ~ %
\eqref{rao}, the perturbed Einstein field equation can be obtained:

\begin{equation}
\begin{aligned} & 6H\left(H\Phi+\dot{\Psi}\right)-\frac{2k^{2}}{a^{2}}\\ = &
\kappa^{2}f'f^{-2}\left(1-\frac{1}{\sqrt{1+f\dot{\varphi}^{2}+K
fV}}\right)\delta\varphi+\kappa^{2}\frac{f'\dot{\varphi}^{2}\delta\varphi+f
\left(\dot{\varphi}\delta\dot{\varphi}+\dot{\varphi}^{2}\Phi\right)+K
\left(f'V+fV'\right)\delta\varphi}{\left(1+f\dot{\varphi}^{2}+K
fV\right)^{\frac{3}{2}}} \end{aligned}
\end{equation}

\begin{equation}
\begin{aligned} &
2\ddot{\Psi}+6H\left(H\Phi+\dot{\Psi}\right)+2H\dot{\Phi}+4\dot{H\Phi+
\frac{2}{3a^{2}}k^{2}}\left(\Phi-\Psi\right)\\ = &
-\kappa^{2}f'f^{-2}\left(1-\frac{1}{\sqrt{1+f\dot{\varphi}^{2}+K
fV}}\right)\delta\varphi-\kappa^{2}\frac{f'\dot{\varphi}^{2}\delta\varphi+f
\left(\dot{\varphi}\delta\dot{\varphi}+\dot{\varphi}^{2}\Phi\right)+K
\left(f'V+fV'\right)\delta\varphi}{\left(1+f\dot{\varphi}^{2}+K
fV\right)^{\frac{3}{2}}} \end{aligned}
\end{equation}

\begin{equation}
\begin{aligned} &
2\ddot{\Psi}+6H\left(H\Phi+\dot{\Psi}\right)+2H\dot{\Phi}+4\dot{H\Phi+
\frac{2}{3a^{2}}k^{2}}\left(\Phi-\Psi\right)\\ &
=-\kappa^{2}f'f^{-2}\left(1-\frac{1}{\sqrt{1+f\dot{\varphi}^{2}+K
fV}}\right)\delta\varphi-\kappa^{2}\frac{f'\dot{\varphi}^{2}\delta\varphi+f
\left(\dot{\varphi}\delta\dot{\varphi}+\dot{\varphi}^{2}\Phi\right)+K
\left(f'V+fV'\right)\delta\varphi}{\left(1+f\dot{\varphi}^{2}+K
fV\right)^{\frac{3}{2}}} \end{aligned}
\end{equation}

\begin{equation}
H\Phi+\dot{\Psi}=-\frac{\kappa^{2}H^{2}}{\sqrt{1+f\dot{\varphi}^{2}+K fV}}
\dot{\varphi}\delta\varphi
\end{equation}

\begin{equation}
\Phi-\Psi=0
\end{equation}

Performing variation on the equation of motion ~\eqref{eq:zuoyongliang} for $%
\varphi $, we get:

\begin{equation}
\begin{aligned} & \frac{\delta\ddot{\varphi}}{1+f\dot{\varphi}^{2}+K
fV}-\ddot{\varphi}\frac{f'\dot{\varphi}^{2}\delta\varphi+f\left(\dot{
\varphi}\delta\dot{\varphi}+\dot{\varphi}^{2}\Phi\right)+K\left(f'V+fV'
\right)\delta\varphi}{\left(1+f\dot{\varphi}^{2}+K fV\right)^{2}}\\ &
+\left(f''f^{-2}-2f'^{2}f^{-3}\right)\left(1+f\dot{\varphi}^{2}+\frac{1}{2}K
fV+\sqrt{1+f\dot{\varphi}^{2}+K fV}\right)\delta\varphi\\ &
+3H\delta\dot{\varphi}+\dot{\varphi}\left(\dot{\Phi}+3\dot{\Psi}\right)-K
f'f^{-2}\left(-f'V+fV'\right)\delta\varphi-\frac{1}{2}f'f^{-1}\left(\dot{
\varphi}\delta\dot{\varphi}+\dot{\varphi}^{2}\Phi\right)\\ &
+\frac{f'^{2}f^{-2}\dot{\varphi}^{2}\delta\varphi+f'f^{-1}\left(\dot{
\varphi}\delta\dot{\varphi}+\dot{\varphi}^{2}\Phi\right)+K
f'f^{-2}\left(-f'V+fV'\right)\delta\varphi}{2\sqrt{1+f\dot{\varphi}^{2}+K
fV}}-\frac{1}{2}K V''\delta\varphi\\ & =0 \end{aligned}
\end{equation}

In order to obtain scalar perturbations and tensor perturbations in our
model, it is sufficient to consider the slow-rolling approximation of
large-scale $k\ll aH$. In this case $\ddot{\varPhi}$,$\ddot{\Psi}$,$\dot{ %
\varPhi}$,$\dot{\Psi}$ can be ignored (see ~\cite%
{Amendola:2005cr,Amendola:2007ni,Nozari:2012cy,Nozari:2013mba}). Therefore,
in a large range, the perturbation equation of motion can be written in the
following form:

\begin{equation}
3H\delta\dot{\varphi}+\mathcal{K}\delta\varphi=0  \label{raodong}
\end{equation}

\noindent and

\begin{equation}
\dot{\varphi}\delta\dot{\varphi}+\dot{\varphi}^{2}\Phi=0
\end{equation}

\noindent where

\begin{equation}
\begin{aligned} \mathcal{K}= &
\left(f''f^{-2}-2f'^{2}f^{-3}\right)\left(1+\frac{1}{2}K fV+\sqrt{1+K
fV}\right)-K f'f^{-2}\left(f'V+fV'\right)\left(1-\frac{1}{2}\left(1+K
fV\right)^{-\frac{1}{2}}\right)\\ & -\frac{1}{2}K f'^{2}f^{-2}V+\frac{1}{2}K
f''f^{-1}V+\frac{1}{2}K f^{-1}f'V'+\frac{1}{2}K V''\end{aligned}
\end{equation}

For small scale perturbations, since the spatial distance is very small, the
space can be regarded as flat at this time. At this time, the perturbation
should be a quantum field theory plane wave solution in a flat spacetime.
Therefore, the perturbation of the co-moving curvature on the super-horizon
scale is almost constant:

\begin{equation}
d\ln k\left(\varphi\right)=dN\left(\varphi\right)
\end{equation}

Therefore, the perturbation is decomposed into two parts, one part is
parallel to the cosmic brane trajectory, called adiabatic perturbation or
curvature perturbation (if there is only one scalar field during inflation,
we will deal with this type of perturbation~\cite%
{Bassett:1999mt,Gordon:2000hv,Kaloper:2005wa,Maartens:1999hf}); the other
part is the perturbation orthogonal to the trajectory, which we call entropy
perturbation or scalar perturbation ~\cite%
{Bassett:1999mt,Gordon:2000hv,Langlois:2000ns,Langlois:2006vv}.

In order to obtain the explicitness of the perturbation field $\delta\varphi$%
, we introduce the function $\mathcal{A}$:

\begin{equation}
\mathcal{A}=\frac{V^{\prime }}{V}\delta\phi
\end{equation}

Then, we can rewrite the eq.~\eqref{raodong} as:

\begin{equation}
\frac{\mathcal{A}^{\prime }}{\mathcal{A}}=\frac{V^{\prime }}{V}-\frac{
V^{\prime \prime }}{V^{\prime }}+\frac{\mathcal{K}}{-f^{\prime -2}\sqrt{1+K
fV}+f^{\prime -2}+\frac{1}{2}K f^{\prime -1}V-\frac{1}{2}K V^{\prime }}
\end{equation}

We can get:

\begin{equation}
\mathcal{A}=C\exp\left(\intop\frac{\mathcal{A}^{\prime }}{\mathcal{A}}
\delta\phi\right)
\end{equation}

\noindent where $C$ is the integral constant.

So we can solve the perturbation field $\delta\phi$

\begin{equation}
\delta\phi=\frac{CV^{\prime }}{V}\exp\left(\int\left[\frac{V^{\prime }}{V}-
\frac{V^{\prime \prime }}{V^{\prime }}+\frac{\mathcal{K}}{-f^{\prime -2}
\sqrt{1+K fV}+f^{\prime -2}+\frac{1}{2}K f^{\prime -1}V-\frac{1}{2}K
V^{\prime }}\right]d\phi\right)
\end{equation}

We can get the following expressions for the density perturbation amplitude:

\begin{equation}
A_{s}^{2}=\frac{k^{3}C}{2\pi^{2}}\frac{V^{\prime 2}}{V^{2}}\exp\left(\int %
\left[\frac{V^{\prime }}{V}-\frac{V^{\prime \prime }}{V^{\prime }}+\frac{
\mathcal{K}}{-f^{\prime -2}\sqrt{1+K fV}+f^{\prime -2}+\frac{1}{2}K
f^{\prime -1}V-\frac{1}{2}K V^{\prime }}\right]d\varphi\right)
\end{equation}

Further obtain the spectrum index of the original scalar perturbation:

\begin{equation}
n_{s}-1=\frac{d\ln A_{s}^{2}}{dN\left(\varphi\right)}=\frac{1}{3H^{2}}\left[
\frac{V^{\prime }}{V}-\frac{V^{\prime \prime }}{V^{\prime }}+\frac{\mathcal{%
K }}{-f^{\prime -2}\sqrt{1+K fV}+f^{\prime -2}+\frac{1}{2}K f^{\prime -1}V-
\frac{1}{2}K V^{\prime }}\right]
\end{equation}

Similarly, when away from the Hubble radius, the tensor perturbation
amplitude in this mode is:

\begin{equation}
A_{T}^{2}=\frac{4\kappa^{2}}{25\pi}H^{2}|_{k=aH}=\frac{4\kappa^{4}}{75}\frac{
1}{f}\left(\sqrt{1+K fV}-1\right)
\end{equation}

The ratio between the tensor perturbation amplitude and the scalar
perturbation amplitude (tensor-scalar ratio $r$) is another important
parameter , which is given by:

\begin{equation}
\begin{aligned}r= &
\frac{A_{T}^{2}}{A_{s}^{2}}=\frac{8\kappa^{4}}{75k^{3}C}\frac{V^{2}\left(
\sqrt{1+K fV}-1\right)}{V'^{2}f}\\ &
\times\exp\left(-\int\left[\frac{V'}{V}-\frac{V''}{V'}+\frac{
\mathcal{K}}{-f'f^{-2}\sqrt{1+K fV}+f'f^{-2}+\frac{1}{2}K
f'f^{-1}V-\frac{1}{2}K V'}\right]d\varphi\right) \end{aligned}  \label{r}
\end{equation}

\section{Constraints of Observation on the Model}

In the cosmological equation of the model, the new general DBI action has
two functions, which play an important role in the dynamics of the model.
They are $f\left( \varphi \right) $ and $V\left( \varphi \right) $. $f\left(
\varphi \right) $ is given by the curl factor. In pure ~$Ads_{5}$~, $f\left(
\varphi \right) $ takes a simple form, namely $f\left( \varphi \right)
=\beta \varphi ^{-4}$. On the other hand, the reference ~\cite%
{Tsujikawa:2013ila} introduced another function of $f\left( \varphi \right) $
as $f\left( \varphi \right) =\beta e^{-\kappa \varphi }$. Therefore, we
divide this section into two subsections. One has $f\left( \varphi \right)
=\beta \varphi ^{-4}$ and the other has $f\left( \varphi \right) =\beta
e^{-\kappa \varphi }$ . Then, we conduct the research by choosing the form
of potential energy.

\subsection{$f\left(\protect\varphi\right)=\protect\beta\protect\varphi^{-4}$%
}

For this type of $f\left(\varphi\right)$, we consider two types of
potential: $V\left(\varphi\right)=\frac{\sigma}{2}\varphi^{2}$ quadratic and
quartic potential $V\left(\varphi\right)=\frac{\sigma}{4}\varphi^{4}$. We
observe that the parameter $K$ can be absorbed into the potential function $V
$. Then, we redefine $\sigma^{\prime }=K\sigma$, then the parameter we have
considered becomes $\left\{ \beta, \sigma^{\prime }\right\}$. Next, we
obtain some constraints on the model parameters by analyzing these
parameters in the context of Planck + WMAP9 + BAO data.

$V\left(\varphi\right)=\frac{\sigma}{2}\varphi^{2}$

We consider the quadratic potential $V\left(\varphi\right)=\frac{\sigma}{2}
\varphi^{2}$.

We introduce $\epsilon\equiv\frac{\dot{H}}{H}=1 $ (corresponding to the end
of the inflation), and we get $\varphi_{f}$. Then, by substituting the
result into $N=\int_{t_{hc}}^{t_{f}}Hdt$, we get $\varphi_{hc}$. By
substituting $\varphi_{hc}$ into $n_{s}-1$ and $r=\frac{A_{T}^{2}}{A_{s}^{2}}
$, we draw a relationship between the tensor-scalar ratio and the spectral
index. In the context of Planck + WMAP9 + BAO joint data, the following
figure is drawn for $N=50$ and $N=60$ (Figure ~\ref{b4s2}):

\begin{figure*}[ht]

\centering
\includegraphics[scale=0.5]{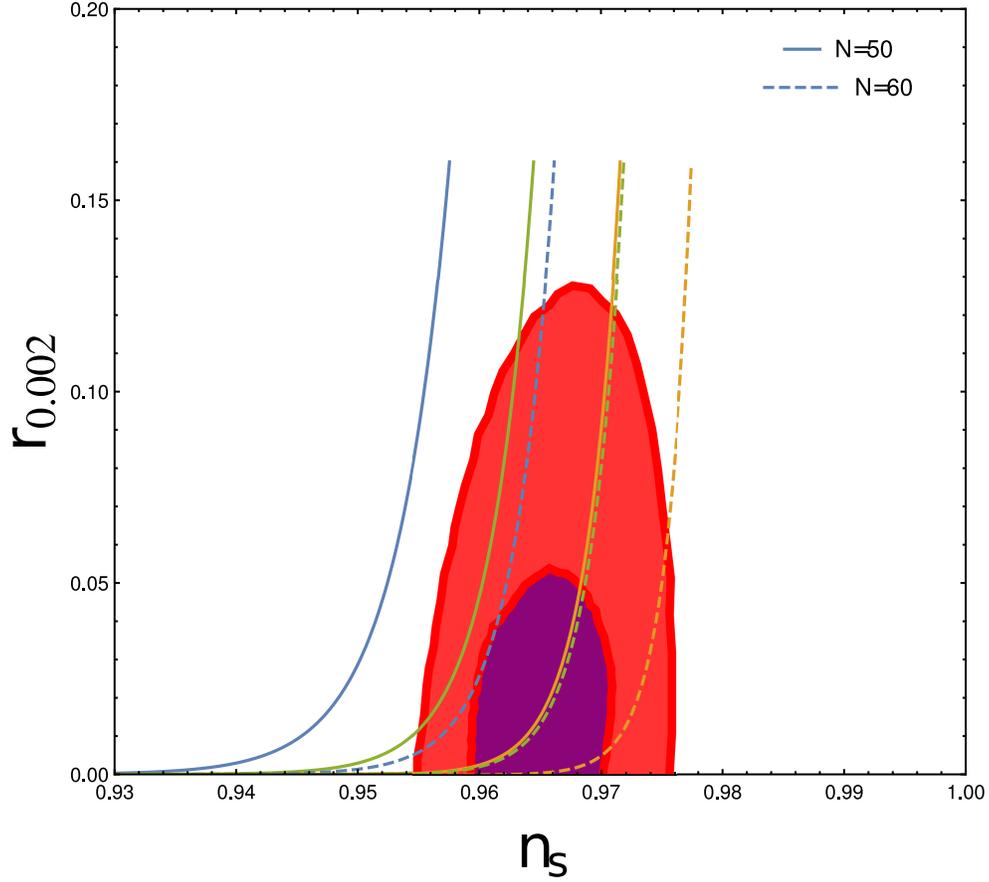}
\caption{Relationships between the tensor-scalar ratio $r$ and the spectral
index  $n_{s}$ for parameters $\left\{ \sigma ^{\prime
},\beta \right\} $ = $\left\{ 0.8835,9.5547\right\} $, $\left\{ 1.1440,0.7592\right\} $, $\left\{
1.4134,3.5523\right\} $}
\label{b4s2}
\end{figure*}

As shown in figure ~\ref{b4s2}, for some parameter combinations of $\left\{
\sigma^{\prime },\beta\right\}$, the model is compatible with Planck + WMAP9
+ BAO data.

In figure~\ref{b4s2} when the parameter group $\left\{ \sigma ^{\prime
},\beta \right\} $ take $\left\{ 0.8835,9.5547\right\} $, the model is at $%
N=50$. It is not compatible with Planck + WMAP9 + BAO data. When $N=60$, the
model can be compatible with 67$\%$ CL of Planck + WMAP9 + BAO data. When
the parameter group $\left\{ \sigma ^{\prime },\beta \right\} $ in the
figure is taken as $\left\{ 1.1440,0.7592\right\} $, when $N=50$, the model
can match Planck + WMAP9 + The 67$\%$ CL of the BAO data, which is
compatible. When $N=60$, the model can be compatible with the 95$\%$ CL of
the Planck + WMAP9 + BAO data. When the parameter group $\left\{ \sigma
^{\prime },\beta \right\} $ in the figure is taken as $\left\{
1.4134,3.5523\right\} $, when $N=50$, the model can match Planck + WMAP9 +
The 95$\%$ CL of the BAO data, which is compatible. When $N=60$, the model
can be compatible with the 95$\%$ CL of the Planck + WMAP9 + BAO data.

Figure ~\ref{b4s2} shows three curves with different parameters. Since the
curve is from eq.~\eqref{r}:

\begin{equation}
\begin{aligned}r= &
\frac{A_{T}^{2}}{A_{s}^{2}}=\frac{8\kappa^{4}}{75k^{3}C}\frac{V^{2}\left(
\sqrt{1+K fV}-1\right)}{V'^{2}f}\\ &
\times\exp\left(-\int\left[\frac{V'}{V}-\frac{V''}{V'}+\frac{
\mathcal{K}}{-f'f^{-2}\sqrt{1+K fV}+f'f^{-2}+\frac{1}{2}K
f'f^{-1}V-\frac{1}{2}K V'}\right]d\varphi\right) \end{aligned}
\end{equation}

\noindent i.e.

\begin{equation}
r=\frac{8\kappa^{4}}{75k^{3}C}\frac{V^{2}\sqrt{1+fV}}{V^{\prime 2}}
\exp\left(-3\int H^{2}\left(n_{s}-1\right)d\varphi\right)
\end{equation}

Therefore, we can see that $r$ and $n_{s}$ are exponentially related.

\subsection{$f\left(\protect\varphi\right)=\protect\beta e^{-\protect\kappa\protect\varphi}$}

For this type of $f\left(\varphi\right)$, we consider two types of
potential: $V\left(\varphi\right)=\frac{\sigma}{2}\varphi^{2}$ quadratic
potential, fourth-order potential $V\left(\varphi\right)=\frac{\sigma}{4}%
\varphi^{4}$. Next, we obtain some constraints on the model parameters by
analyzing these parameters in the context of Planck + WMAP9 + BAO data.

A:$V\left(\varphi\right)=\frac{\sigma}{2}\varphi^{2}$

In the first step, we consider the quadratic potential $V\left(\varphi%
\right)=\frac{\sigma}{2}\varphi^{2}$.

We introduce $\epsilon\equiv\frac{\dot{H}}{H}=1 $ (corresponding to the end
of the inflation), and we get $\varphi_{f}$. Then, by substituting the
result into $N=\int_{t_{hc}}^{t_{f}}Hdt$, we get $\varphi_{hc}$. By
substituting $\varphi_{hc}$ into $n_{s}-1$ and $r=\frac{A_{T}^{2}}{A_{s}^{2}}
$, we draw a relationship between the tensor-scalar ratio and the spectral
index. In the context of Planck + WMAP9 + BAO joint data, the following
figure is drawn for $N=50$ and $N=60$ (Figure ~\ref{bes2}).

\begin{figure*}[ht]

\centering
\includegraphics[scale=0.5]{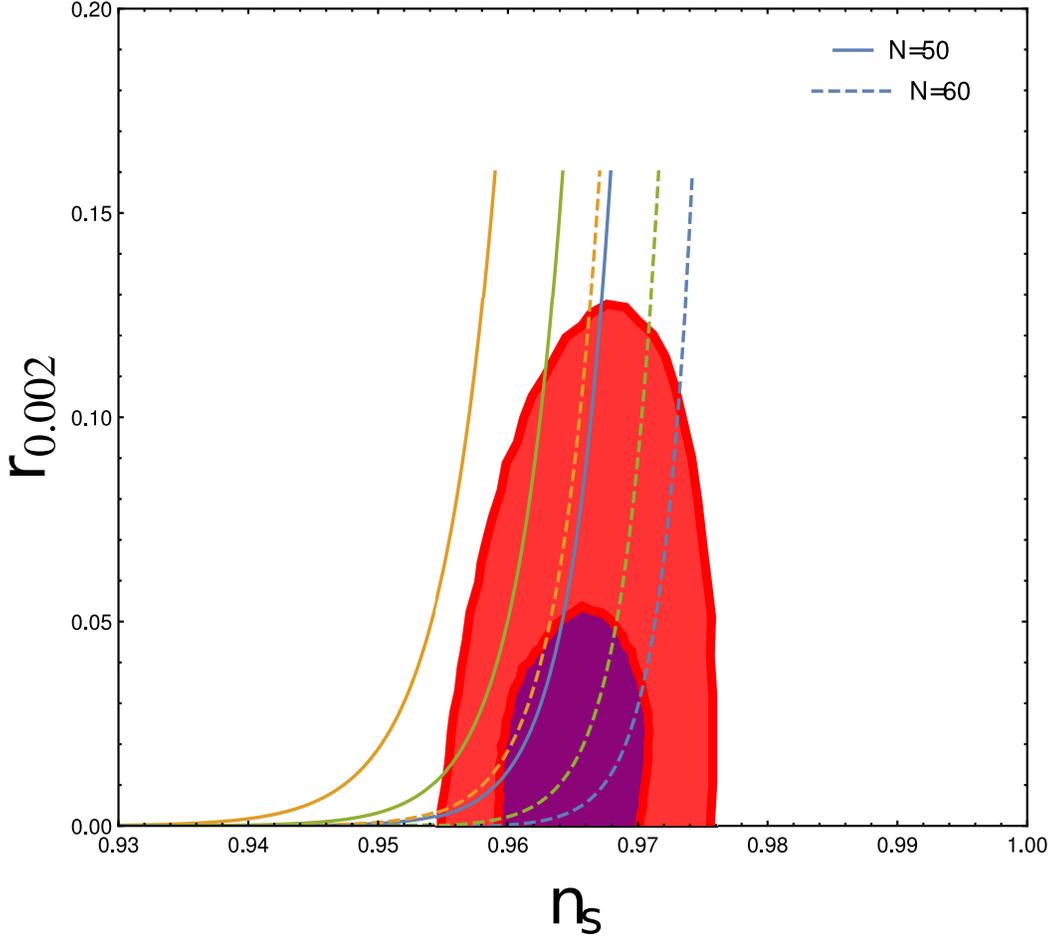}
\caption{Relationships between the tensor-scalar ratio $r$ and the spectral
index  $n_{s}$ for parameters $\left\{ \sigma ^{\prime
},\beta \right\} $ =$\left\{ 1.1405, 7.3919\right\}$, $%
\left\{ 1.1141, 1.0673\right\}$, $\left\{ 0.8770, 8.4078\right\}$}
\label{bes2}
\end{figure*}

In the figure ~\ref{bes2}, parameter group $\left\{ \sigma^{\prime
},\beta\right\}$ takes $\left\{ 1.1405, 7.3919\right\}$, the model is at $%
N=50$. It is not compatible with Planck + WMAP9 + BAO data. When $N=60$, the
model can be compatible with 95$\%$ CL of Planck + WMAP9 + BAO data. When
the parameter group $\left\{ \sigma^{\prime },\beta\right\}$ is taken as $%
\left\{ 1.1141, 1.0673\right\}$ in the figure, that the model can be
compared with Planck + WMAP9 + The 67$\%$ CL of the BAO data is compatible.
When $N=60$, that the model can be compatible with the 95$\%$ CL of the
Planck + WMAP9 + BAO data. When the parameter group $\left\{ \sigma^{\prime
},\beta\right\}$ in the figure is taken as $\left\{ 0.8770, 8.4078\right\}$,
when $N=50$, that the model can match Planck + WMAP9 + The 95$\%$ CL of the
BAO data is compatible. When $N=60$, the model can be compatible with the 95$%
\%$ CL of the Planck + WMAP9 + BAO data.

The figure ~\ref{bes2} shows three curves with different parameters. Since
the curve is from eq.~\eqref{r}:

\begin{equation}
\begin{aligned}r= &
\frac{A_{T}^{2}}{A_{s}^{2}}=\frac{8\kappa^{4}}{75k^{3}C}\frac{V^{2}\left(%
\sqrt{1+K fV}-1\right)}{V'^{2}f}\\ &
\times\exp\left(-\int\left[\frac{V'}{V}-\frac{V''}{V'}+\frac{%
\mathcal{K}}{-f'f^{-2}\sqrt{1+K fV}+f'f^{-2}+\frac{1}{2}K
f'f^{-1}V-\frac{1}{2}K V'}\right]d\varphi\right) \end{aligned}
\end{equation}

\noindent i.e.

\begin{equation}
r=\frac{8\kappa^{4}}{75k^{3}C}\frac{V^{2}\sqrt{1+fV}}{V^{\prime 2}}%
\exp\left(-3\int H^{2}\left(n_{s}-1\right)d\varphi\right)
\end{equation}

Therefore, we can see that $r$ and $n_{s}$ are exponentially related.

B:$V\left(\varphi\right)=\frac{\sigma}{4}\varphi^{4}$

In the second step, we consider the quartic potential $V\left(\varphi\right)=%
\frac{\sigma}{4}\varphi^{4}$.

We introduce $\epsilon\equiv\frac{\dot{H}}{H}=1$ (corresponding to the end
of the inflation), and we get $\varphi_{f}$. Then, by substituting the
result into $N=\int_{t_{hc}}^{t_{f}}Hdt$, we get $\varphi_{hc}$. By
substituting $\varphi_{hc}$ into $n_{s}-1$ and $r=\frac{A_{T}^{2}}{A_{s}^{2}}
$, we draw a relationship between the tensor-scalar ratio and the spectral
index. In the context of Planck + WMAP9 + BAO joint data, the following
figure is drawn for $N=50$ and $N=60$ (Figure ~\ref{bes4}).

\begin{figure*}[ht]

\centering
\includegraphics[scale=0.5]{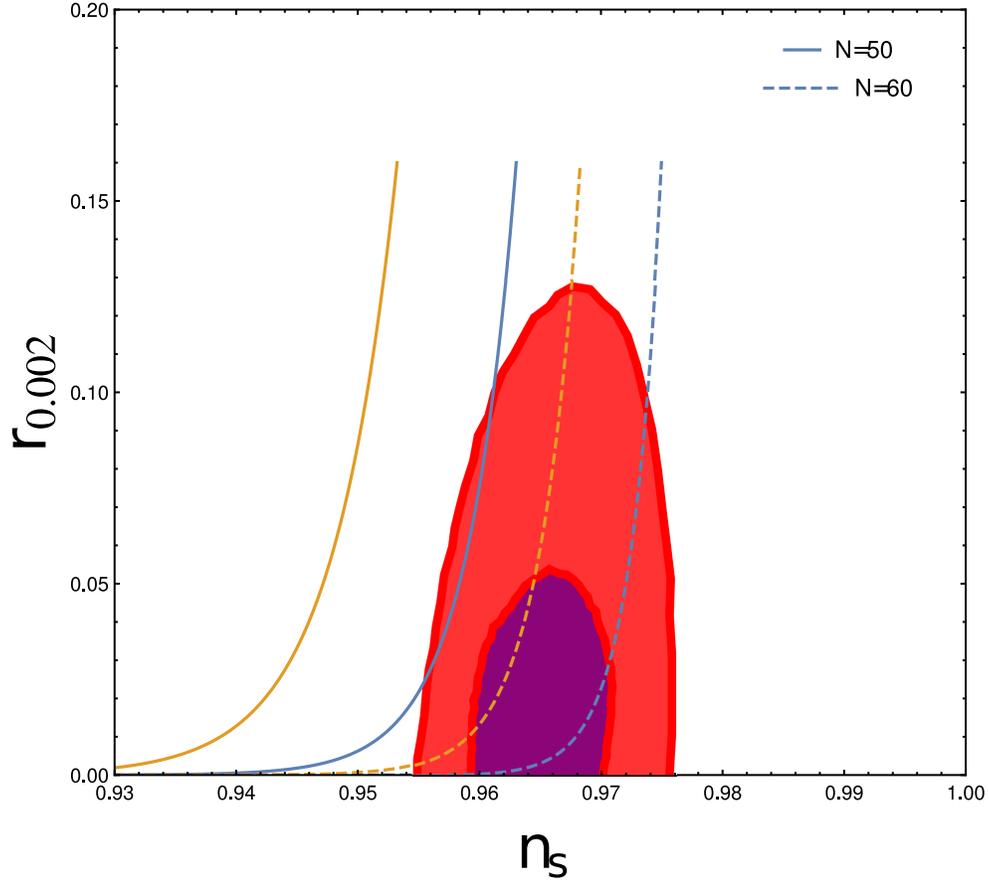}
\caption{Relationships between the tensor-scalar ratio $r$ and the spectral
index  $n_{s}$ for parameters $\left\{ \sigma^{\prime
},\beta\right\}$ = $\left\{ 0.0023,3.5974\right\}$, $%
\left\{ 0.0029,7.7111\right\}$}
\label{bes4}
\end{figure*}

In the figure ~\ref{bes4}, parameter group $\left\{ \sigma^{\prime
},\beta\right\}$ takes $\left\{ 0.0023,3.5974\right\}$, the model is at $N=50
$ ,it is not compatible with Planck + WMAP9 + BAO data. When $N=60$, the
model can be compatible with 95$\%$ CL of Planck + WMAP9 + BAO data. When
the parameter group $\left\{ \sigma^{\prime },\beta\right\}$ is taken as $%
\left\{ 0.0029,7.7111\right\}$ in the figure, when $N=50$, that the model
can match Planck + WMAP9 + 67$\%$ CL of BAO data is compatible, and the
model can be compatible with Planck + WMAP9 + 95$\%$ CL of BAO data when $%
N=60$.

The figure ~\ref{bes4} shows two curves with different parameters. Since the
curve is from eq.~\eqref{r}:

\begin{equation}
\begin{aligned}r= &
\frac{A_{T}^{2}}{A_{s}^{2}}=\frac{8\kappa^{4}}{75k^{3}C}\frac{V^{2}\left(%
\sqrt{1+K fV}-1\right)}{V'^{2}f}\\ &
\times\exp\left(-\int\left[\frac{V'}{V}-\frac{V''}{V'}+\frac{%
\mathcal{K}}{-f'f^{-2}\sqrt{1+K fV}+f'f^{-2}+\frac{1}{2}K
f'f^{-1}V-\frac{1}{2}K V'}\right]d\varphi\right) \end{aligned}
\end{equation}

\noindent i.e.

\begin{equation}
r=\frac{8\kappa^{4}}{75k^{3}C}\frac{V^{2}\sqrt{1+fV}}{V^{\prime 2}}%
\exp\left(-3\int H^{2}\left(n_{s}-1\right)d\varphi\right)
\end{equation}

Therefore, we can see that $r$ and $n_{s}-1$ are exponentially related.

\section{Conclusion}

In this article, we deduce a new general DBI action in equal rights for
kinetic energy and potential energy in string theory, and introduce it into early cosmology,
and calculate the inflation parameters and original density perturbations in
detail. We deduce the influence of the scalar itself on the amplitude of the
density perturbation, so that when different curl factors and potential
functions are introduced, we can see the consistent results of the scalar
field participating in the process of forming the perturbation. We see that
because the scalar field is local, the tensor-scalar ratio is approximate to
an exponential function. The parameters $\left\{ \sigma ^{\prime },\beta
\right\} $ brought by the curl factor and the potential function affect the
shape of the exponential function, thereby we find that these lines relative to the parameters pass through
the range defined by the current observation data, which means that the theory of this paper conforms to the current  experiments.

Therefore, the new general DBI action is important in string theory, produces the inflation, and it
yields enough original density perturbations to form the universe we see
today. The new general DBI action solve the current general DBI action
paradox of the conversion of kinetic energy and potential energy in equal rights in string theory, and it
predicts the power spectrum of satisfying the current observations, thus
showing that the theory of this paper is consistent, supplementing the existing
theories, and also a new way of explaining the inflation of the universe.

\begin{acknowledgments}
The work is supported by National Natural Science Foundation of China (No.
11875081).
\end{acknowledgments}

\end{document}